Full characterization of few-cycle superradiant pulses generated from a free-electron laser oscillator

Heishun Zen[1*], Ryoichi Hajima[2], and Hideaki Ohgaki[1]

[1]Institute of Advanced Energy, Kyoto University, Gokasho Uji, 611-0011 Kyoto, Japan
[2]National Institutes for Quantum Science and Technology, Kizugawa, Kyoto 619-0215, Japan
*e-mail: zen@iae.kyoto-u.ac.jp

**Abstract**
The detailed structure of few-cycle superradiant pulses generated from a free-electron laser (FEL) oscillator was experimentally revealed for the first time. As predicted in numerical simulations, the obtained pulse structures have ringing, i.e., a train of subpulses following the main pulse. Owing to the phase retrieval with a combination of linear and nonlinear autocorrelation measurements, the temporal phase was also obtained with the pulse shape, and π-phase jumps between each pulse were confirmed. This particular pulse shape with π-phase jumps is a common feature of Burnham-Chiao ringing or superradiance ringing. Using unaveraged FEL simulations, the ringing was found to originate from repeated formation and deformation of microbunches across successive bunch slices. The simulation results imply that there should be superluminal propagation of the FEL pulse due to strong gain modulation along the undulator of the FEL oscillator.

**Main**
Superradiance (SR), originally predicted by Dicke in 1954 [1], is a quantum optical process in an N-body system coupled by a common electromagnetic field, in which the system exhibits accelerated radiative decay, generating a pulse of coherent light whose peak intensity is $\propto N^2$ and pulse duration is $\propto 1/N$. Since Dicke's first paper, many theoretical and experimental works have been conducted on SR in gases [2, 3, 4, 5, 6, 7, 8] and solids [9, 10, 11, 12].

The first observation of SR was reported with optically pumped HF gas at a wavelength of 84 μm [2]. In the experiment, ringing following the main pulse was observed in the SR pulse in addition to confirmation of the $N^2$ dependence of the peak intensity. The ringing was later attributed to coherent Rabi-type oscillations similar to those predicted by Burnham and Chiao [13, 14]; thus, the phenomenon is termed Burnham-Chiao ringing (BC ringing). BC ringing is a visible manifestation of the light-matter resonant interaction and is derived as a solution of the optical Bloch equations, which are the basis for the study of the light-matter resonant interaction to describe SR, photon echo, self-induced transparency, optical free induction decay, and more [15].

In the study of a free-electron laser (FEL), Bonifacio and Casagrande first predicted in 1985 that a single-pass FEL can be operated in the SR regime, where the radiation power is proportional to the square of the number of electrons [16]. Such SR pulses were examined in a single-pass FEL seeded by a Ti:sapphire laser by using a frequency-resolved optical gating technique [17]. In the experiment, nonlinear pulse shortening and formation of weak ringing were observed. The SPARC and FERMI@elettra groups have reported SR pulses generated from a high-gain harmonic-generation (HGHG) FEL [18] and a cascaded HGHG FEL [19, 20]. Theoretical studies revealed that SR can also occur in FEL oscillators [21, 22, 23]. The first demonstration of SR in an FEL oscillator was reported by Jaroszynski et al. [24]. They confirmed the $N^2$ dependence of the peak power in an FEL oscillator, FELIX. There have been several analytical and numerical studies on SR in FEL oscillators to discuss few-cycle pulse generation with BC ringing [23, 25, 26, 27]. Measurements of SR pulses in FEL oscillators have also been conducted but are limited to interpretation of autocorrelation waveforms without phase recovery [24, 28, 29, 30] or to reconstruction of a pulse with coarse resolution [31].

Here, we report the first full characterization of few-cycle SR pulses generated from an FEL oscillator. The structure of the FEL pulses was precisely evaluated via a fringe-resolved autocorrelation measurement and a phase retrieval analysis. As a result, an SR pulse containing multiple subpulses and π-phase jumps between these pulses were observed. Those are clear evidence of BC ringing. The observed pulse structures, i.e., the intensity and the phase, were well reproduced by one-dimensional numerical simulations. The numerical results indicate that the ringing with π-phase jumps originates from repeated formation and deformation of microbunches across successive bunch slices and superluminal evolution of the pulse in the FEL oscillator.

## Results

The experiments were carried out at KU-FEL, a mid-infrared FEL facility at Kyoto University [32]. KU-FEL has two operation modes of the electron gun: thermionic cathode operation and photocathode operation. For the lasing wavelength of approximately 11 μm, the efficiency of energy extraction from the kinetic energy of the electron beam to the FEL at the end of the macropulse was 5.5% and 9.4% for thermionic and photocathode operation, respectively [33, 34]. Under these two operation modes, the temporal structure of FEL pulses was precisely retrieved from simultaneously obtained linear autocorrelation (LA) and nonlinear fringe-resolved autocorrelation (FRAC) traces. While FRAC is widely used for optical pulse measurements, it does not directly provide the optical phase information to reconstruct the pulse shape. For precise reconstruction of the FEL pulse structure, i.e., recovery of the intensity and phase, we adopted the "Evolutionary Phase Retrieval from Interferometric AutoCorrelation (EPRIAC)" algorithm [35], in which the frequency spectrum of FEL pulses deduced by Fourier transformation of the LA trace is used along with the FRAC trace for the reconstruction.

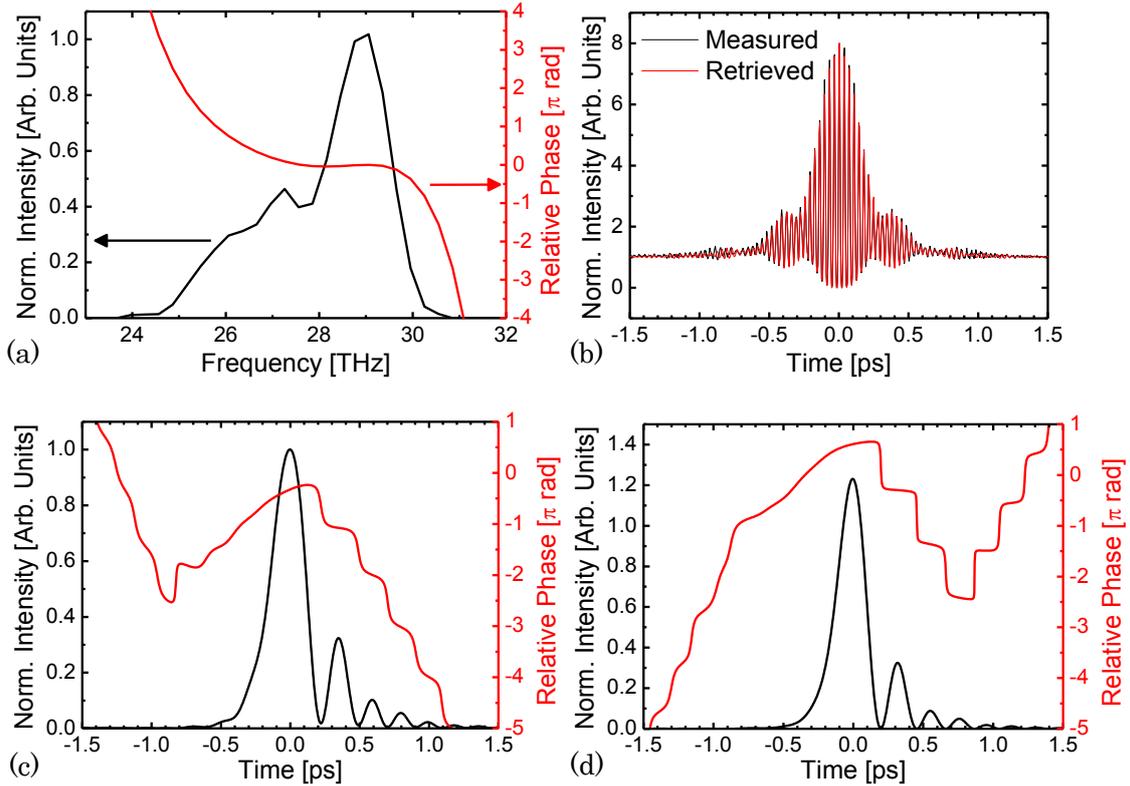

Figure 1: (a) FEL spectrum deduced by Fourier transformation of the LA trace and retrieved relative spectral phase. (b) Measured and retrieved FRAC traces. (c) Retrieved pulse structure and relative phase. (d) Retrieved pulse structure and relative phase after subtraction of the dispersion effect of the refractive optical components in the beam path.

*Pulse structure and phase of the FEL pulse under the thermionic mode*

Figure 1 shows the pulse measurement results for the average of a 270 ns time window around the peak of the FEL macropulse in the thermionic cathode mode with an extraction efficiency of 5%. In the measurement, an FEL pulse at the end of a macropulse was obtained by gating the signals. Figure 1(a) shows the frequency spectrum deduced from an LA trace and the spectral phase retrieved by EPRIAC. The measured and retrieved FRAC traces are shown in Figure 1(b). The measured FRAC traces can be well reproduced by the retrieved phase. The retrieved pulse shape and the relative temporal phase are shown in Figure 1(c). We removed the direction-of-time ambiguity in EPRIAC by observing the pulse elongation when a 10-mm-thick ZnSe window with a group velocity dispersion (GVD) of -1365.9 $fs^2$/mm at 10.3 μm was inserted into the beam path. The details of the pulse-direction determination are explained in Supplementary Information S2.

Into the FEL transport line from the out-coupling hole of the FEL optical cavity to the autocorrelation apparatus, two KRS-5 windows with a total thickness of 7 mm and one KBr window with a thickness of 3 mm were inserted. In addition, the autocorrelation apparatus had a ZnSe beam splitter with an effective thickness of 3.1 mm, and each beam of the autocorrelator passed through it once before hitting the second harmonic generation (SHG) crystal. In the few-cycle pulse measurements, the group delay dispersion (GDD) of these refractive optical components cannot be neglected. Since the relative phase distribution of the FEL pulse in the frequency domain was retrieved, the influence of these refractive optical components can be subtracted from the retrieved phase, and we can obtain the precise structure of the FEL pulse immediately after the out-coupling hole of the FEL optical cavity. The details of the subtraction of the influence of the refractive optical components are described in Supplementary Information S3. The reconstructed pulse shape and phase immediately after the out-coupling hole are shown in Figure 1(d). The pulse duration of the main pulse was approximately 230 fs in full width at half maximum (FWHM), which is equivalent to 6.7 optical cycles of 10.3-μm radiation. The relative phase has sudden π-phase jumps between pulses.

*Pulse structure and phase of the FEL pulse under the photocathode mode*

With the same procedure, the FEL pulse under photocathode operation with an energy extraction efficiency of approximately 9% was retrieved. The evaluated pulse shape and phase of the FEL pulse at the out-coupling hole are shown in Figure 2(b) together with the measured frequency spectrum and the retrieved phase distribution at the out-coupling hole (Figure 2(a)). The pulse duration of the main pulse was approximately 150 fs in FWHM and is equivalent to 4.2 optical cycles of 10.7-µm radiation (peak frequency of 28 THz). In this result, π-phase jumps were also observed. The FEL spectrum shown in Figure 2(a) spans from 20 to 30 THz, which is much wider than that for the thermionic operation case (24.5 to 30 THz, shown in Figure 1(a)).

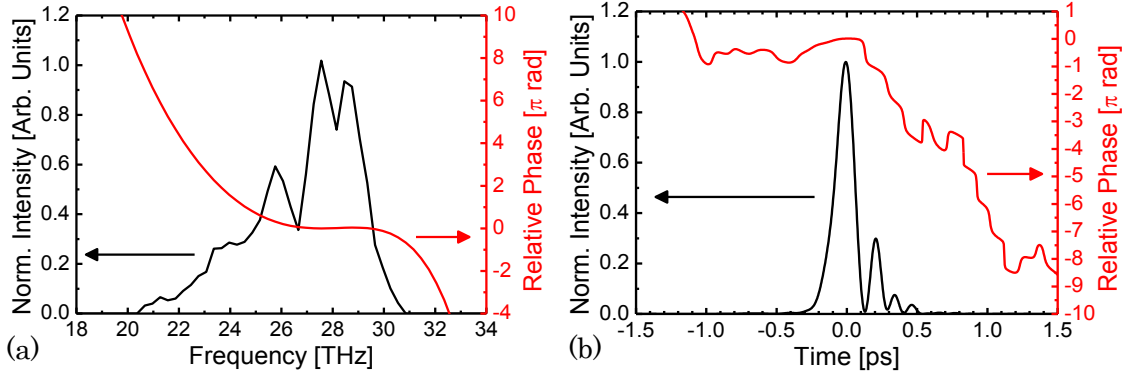

Figure 2: (a) Measured FEL spectrum and spectral phase. (b) Retrieved pulse shape and temporal phase of the FEL pulse generated at KU-FEL under the condition having an energy extraction efficiency of approximately 9%.

*Comparison with numerical simulation*

In Figure 3, the measured pulse structures and phase are plotted together with those obtained by numerical simulations. In both thermionic and photocathode operation, the intensity and phase of the main pulse obtained in the experiments are well reproduced by the numerical simulations. In the case of thermionic operation, the intensity and phase of the subpulses also show good agreement between the experimental and numerical results. In the case of photocathode operation, the numerical results have some discrepancies in the structure of the subpulses. The reasons for the discrepancies are discussed in the following section. Note that compensation of the influence of the refractive optical components in the beam path and the autocorrelation apparatus, which requires information on the spectral phase distribution, is indispensable for comparing the numerical results and experimental results.

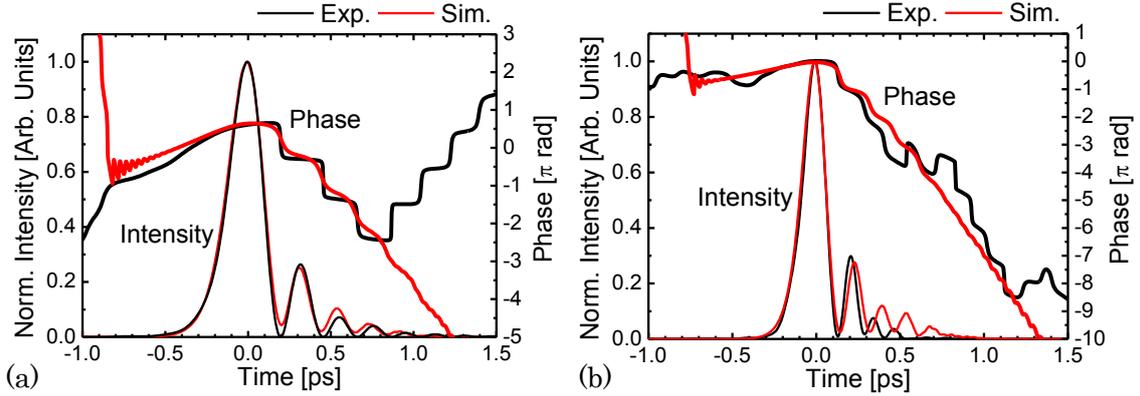

Figure 3: Comparison with 1D numerical simulation results. (a) Thermionic operation. (b) Photocathode operation.

Discussion

The formation of subpulses and $\pi$-phase jumps in the FEL pulses are similar to the ringing observed in the light-matter resonant interaction, Burnham-Chiao ringing or SR ringing [13]. We discuss the mechanism of ringing formation in the FEL pulse with the help of numerical simulations. In Figure 4, we plot the simulated FEL pulse at the end of the macropulse for the photocathode mode together with snapshots of the phase-space distributions. In the plots, the longitudinal variable, $\zeta=(z-ct)/L_s$, is defined in a coordinate system moving at the vacuum speed of light $c$ and normalized by the slippage distance $L_s$, the time for traversing the undulator is normalized by the undulator length, $L_u$, as $\tau=ct/L_u$, and the dimensionless electron energy is defined as $\mu=(\Delta E - E_0)/4\pi N_u$, where $\Delta E$ is the energy deviation from the resonance energy $E_0$ and $N_u$ is the number of undulator periods. Overlaid on the pulse shape, the local intensity changes in a single round trip, including the cavity loss of 3%, are indicated by red dots for amplification and blue dots for reduction. The snapshots represent the macroparticle distribution at different positions in the undulator for a specific bunch slice, whose initial position is $\zeta=0.3$ at $\tau=0$. The green arrows indicate the position of the slice in the FEL pulse when the snapshots were taken. In each snapshot, the electric field of the FEL pulse is plotted with a red line, and local changes in the macroparticle energy are plotted with blue bars. In the snapshots, the number of macroparticles was reduced to improve visibility.

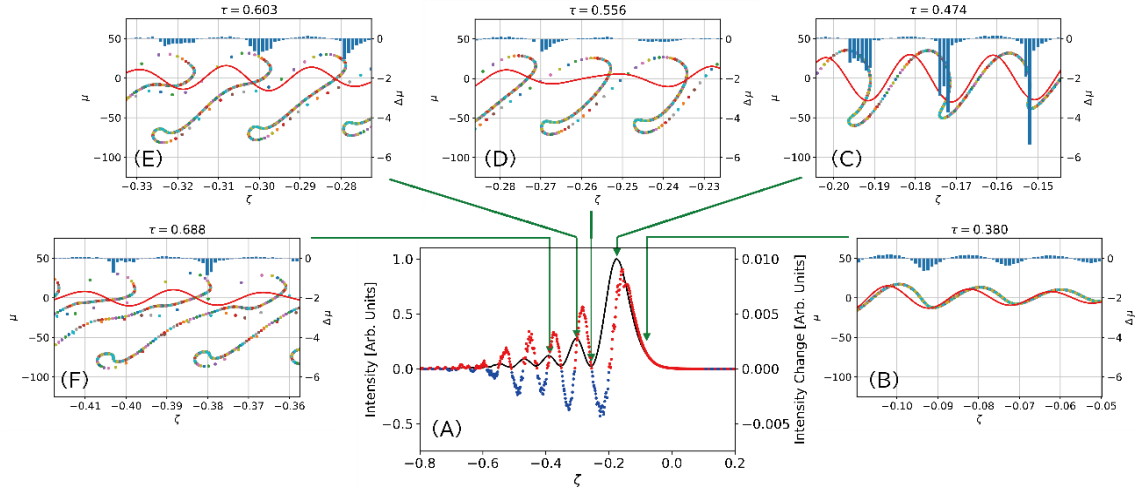

Figure 4: (A) Simulated FEL pulse at the end of the macropulse for the photocathode mode and local intensity changes in a single round trip, including the cavity loss of 3%, with red dots for amplification and blue dots for reduction. (B)-(F) snapshots representing the macroparticle distribution at different positions in the undulator for a specific bunch slice. The red lines in the snapshots indicate the electric field of the FEL pulse, and the blue bars are local changes in the macroparticle energy.

As shown in Figure 4, the radiation at the rising slope of the main pulse is amplified. In contrast, the radiation is reduced due to negative net gain at the falling slope of the first pulse. For the successive subpulses, amplification and reduction occur at the rising and falling slopes, respectively, in a similar manner.

To clarify the origin of the periodic gain modulation, we follow the motion of the macroparticles in the phase space. In the early part of the undulator ($\tau$=0.380, Figure 4(B)), the bunch slices of interest are located at the rising slope of the first pulse, and the FEL electric field induces electron energy modulation. As the slices slip backwards in the FEL pulse, the energy modulation grows into a density modulation, forming microbunches with an interval equal to the FEL wavelength. At the peak of the main pulse ($\tau$=0.474, Figure 4(C)), the electrons in the microbunch lose their energy as they emit strong radiation and thus slip backwards at a faster speed, stretching the phase distribution. At the dip between the main and second pulses ($\tau$=0.556, Figure 4(D)), the density modulation almost disappears, and there is little energy exchange between the electrons and the radiation field. After passing the dip, the bunch slices again induce density modulation, emitting radiation for the second pulse ($\tau$=0.603, Figure 4(E)). Note that the density modulation for the second pulse is established by the overlap of bunch slices over two radiation wavelengths. The third pulse is formed in a similar way by density modulation induced by slices over three radiation wavelengths (τ=0.688, Figure 4(F)). At the trailing edge of each pulse, the envelope of the radiation electric field goes through zero and becomes negative at the leading edge of the next subpulse. This reversal of the sign of the field envelope is the origin of the π-phase jumps. A movie file was prepared to visualize the macroparticle motion in the phase space. See Supplementary Video 1.

Through the above processes, the electrons form microbunches and transfer their energy to the radiation field with high conversion efficiency. Such efficient interaction is possible due to the FEL pulse with a well-designed temporal profile of the intensity and the phase, which is autonomously established after many round trips as an eigenmode of the SR FEL oscillator.

The ringing in the FEL pulse is analogous to the Burnham-Chiao ringing of two-level systems. The BC ringing is interpreted as an oscillation of the state vector on the Bloch sphere, in which phasing and dephasing of the macroscopic polarization of the atoms (or radiation and absorption of the light field by the atoms) occur alternately and the train of radiation pulses has π-phase jumps. The ringing in the FEL pulse originating from the repeated formation and deformation of microbunches is also accompanied by gain modulation and π-phase jumps.

The BC ringing with a π-phase jump in the light-matter resonant interaction is derived by solving the optical Bloch equations [13, 36]. However, there have not been many experimental observations of BC ringing reported to date. Experiments of SR accompanied by BC ringing have been reported for HF gas [2] and quantum dots in perovskite nanocrystals [37], but π-phase jumps have not been examined there. π-phase jumps have been detected in the free-exciton transition in an optically thick semiconductor [38] and the free induction decay in highly absorbing solutions [39], both of which are a type of resonant light-matter interaction, although not SR. BC ringing with π-phase jumps was recently observed for the SR generated from lattice-confined atoms inside a hollow core fibre [40].

π-phase jumps are fundamental properties of SR pulses with BC ringing. Thus, the observation of π-phase jumps for the laser pulses generated from the FEL oscillator is strong evidence that the FEL oscillator operates in the SR regime. Our result is the first full characterization of an SR pulse with a subpicosecond duration.

As a result of the periodic gain modulation shown in Figure 4, the FEL pulse is pushed forwards every round trip, and then, the group velocity of the pulse slightly exceeds the vacuum speed of light. This is known as superluminal propagation, which has been observed in various kinds of laser systems [41, 42, 43, 44]. For FELs, there are theoretical studies on the superluminal propagation in oscillator FELs [27] and seeded single-pass FELs [17, 45]. In the present study, each subpulse is considered to have a superluminal property, and the complex pulse shape is preserved for many round trips due to the periodic modulation of the FEL gain.

The measured widths of the main pulses were 230 and 150 fs for thermionic operation with an extraction efficiency of 5% and photocathode operation with an extraction efficiency of 9%, respectively. The pulse width is almost inversely proportional to the extraction efficiency, as predicted by a theoretical study [23]. This is also evidence that the FEL operates in the SR regime.

There are two possibilities for further shortening of the FEL pulse duration. One is chirp compensation. As shown in Figure 2, the FEL pulse is inherently down-chirped. By compensating for the linear chirp, the FEL pulse duration can be shortened down to 121 fs in FWHM in the case of photocathode operation, which corresponds to 3.4 optical cycles and is 16% longer than the transform-limited pulse (105 fs in FWHM and 2.9 optical cycles). See Supplementary Information S4 for more details. The other possibility is to increase the extraction efficiency by increasing the FEL gain or decreasing the loss of the optical cavity since the pulse duration of the FEL is almost inversely proportional to the efficiency. Experimental validation of these pulse shortening methods is planned at KU-FEL.

The numerical simulation results have discrepancies from the experimental results in the ringing structure. In the simulations, the gaps between the pulses are not zero, and therefore, the phase jumps are blunted. This can be attributed to the limitations of the simulation, such as the numerical diffusion in the finite-difference methods and one-dimensional approximation neglecting the transverse inhomogeneity of the electron and radiation beams. In the case of photocathode operation, the maximum energy decrease of electrons in the experiment was approximately 16%. This large energy change can induce transverse expansion of the electron beam in the latter part of the undulator and disturb the FEL interaction in the subpulses. Three-dimensional FEL simulations with unaveraged codes [46] remain a future work.

## Methods

### Parameters and setup of KU-FEL

The experiments were carried out using the mid-infrared FEL oscillator at the KU-FEL facility [32]. The parameters under thermionic operation and photocathode operation are listed in Table 1.

Table 1: Electron beam and undulator parameters

| Parameters | Thermionic | Photocathode |
|---|---|---|
| Injected electron beam energy | 28.5 MeV | 28.5 MeV |
| Electron bunch charge | 55 pC maximum | ~220 pC |
| Macropulse duration | 6.8 μs | 7 μs |
| Electron bunch repetition rate | 2856 MHz | 29.75 MHz |
| Undulator K-value | 1.35 | 1.35 |
| Undulator period length | 33 mm | 33 mm |
| Number of undulator periods | 52 | 52 |
| Energy extraction efficiency | ~5% | ~9% |

The dynamic cavity desynchronization (DCD) method invented by the FELIX group [47] has been implemented at KU-FEL to obtain high extraction efficiency with a short macropulse electron beam [33]. In each operation mode, DCD parameters, i.e., the timing and amount of frequency jump, were adjusted to obtain the highest output power under these operation modes. The energy extraction efficiencies for the two modes were evaluated from the differences in the electron energy with and without FEL oscillation [33].

*Linear and nonlinear autocorrelation measurement*

A schematic diagram of the autocorrelation setup for simultaneous acquisition of LA and FRAC traces is shown in Figure 5. FEL oscillators inherently generate harmonics due to the higher-order frequency component of microbunched electron beams [48]. The harmonic components were removed by a longpass filter (LPF) with a cutoff wavelength of 7.3 μm (#68-656, Edmund Optics). A pair of off-axis parabolas (OAPs) with reflected focal lengths of 4 inches and 1 inch was used to reduce the FEL beam size by a factor of 4. Two beam splitters (BSs, #62-967, Edmund Optics) were used for splitting and combining the FEL beam. These two BSs, a roof mirror (RM) and two flat mirrors (FMs) were arranged to configure an interferometer. The RM was placed on a linear stage with a minimum step size of 0.5 μm (TSDM40-15X, Sigma-Koki) to change the temporal overlap condition of the split beams. These BSs have a beam-splitting coating and an antireflection coating. The front and rear surfaces of the BSs in the system were arranged as shown in Figure 5. The split FEL beams injected into a nonlinear crystal (NLC) pass through the BS only once to reduce the influence of wavelength dispersion on the FRAC measurement. The split beams reaching detector 1 (DET1, Model 420, ELTEC Instruments, used for measuring the LA) pass through the BS twice, but the influence on the measured spectrum is negligible since the BS has flat spectral properties at the wavelength of interest. Before injection into the NLC, the combined beam was reflected by an OAP with a reflected focal length of 6 inches to make the beam size smaller to increase the SHG intensity. As the NLC, a AgGaSe$_2$ crystal (8×8×0.5 mm, #APO-20-0410-01, 3photon) optimized for SHG at a wavelength of 10 μm was used. The thin crystal enables a wide phase-matching bandwidth. After the NLC, a shortpass filter (SPF) was arranged to block the fundamental component before injection into pyroelectric detector 2 (DET2, Model 420, ELTEC Instruments). DET1 and DET2 were connected to transimpedance amplifiers (TIA60, Thorlabs) to increase the signal intensity. The voltage signal outputs from the amplifiers were connected to an oscilloscope to record the signal intensity. Into the path of the FEL beam from the out-coupling hole of the optical resonator, an optical window made of KRS-5 with a thickness of 3 mm was inserted to reflect a part of the FEL light to monitor the intensity of the FEL by using a pyroelectric detector (DET0, Model 420, ELTEC Instruments) and a transimpedance amplifier (TIA60, Thorlabs). The voltage signal measured by DET1 was divided by the voltage signal measured by DET0. The voltage signal measured by DET2 was divided by the square of the voltage signal measured by DET0 since the SHG intensity depends on the square of the intensity of the injected FEL beam. These pyroelectric detectors combined with the transimpedance amplifiers have a frequency bandwidth of 30 MHz and can resolve the temporal evolution of the FEL macropulse (typically 2 μs and 4 μs in FWHM in the case of thermionic operation and photocathode operation, respectively). To measure the FEL micropulse structure at the end of the electron-beam macropulse, a time gate was arranged at the end of the electron-beam macropulse, and the average of the normalized signals was recorded. At each RM position, the signals of 4 macropulse shots were averaged to reduce the influence of shot-by-shot fluctuation.

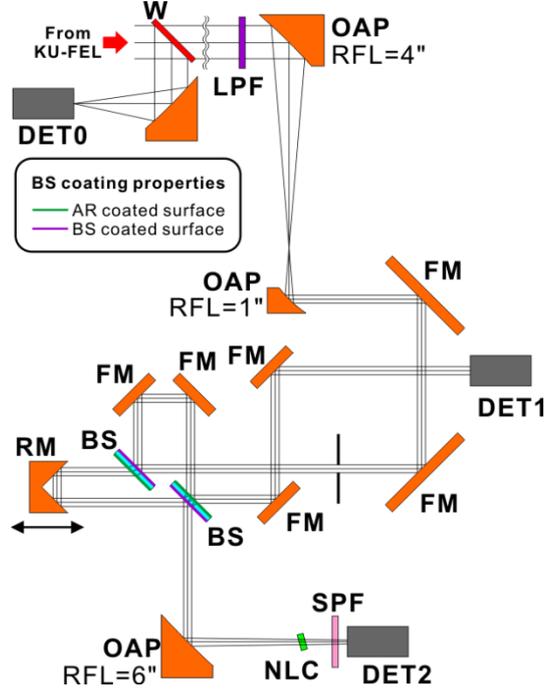

Figure 5: Schematic diagram of the autocorrelation setup for simultaneous acquisition of linear and nonlinear interferometric autocorrelation. LPF: longpass filter, SPF: shortpass filter, NLC: nonlinear crystal for SHG, BS: beam splitter, RM: roof mirror, FM: flat mirror, W: KRS-5 window, DET0: detector for FEL power monitoring, DET1: detector for acquiring LA traces, and DET2: detector for acquiring nonlinear FRAC traces.

*Retrieval of the spectral phase and time structure*

From the measured linear and nonlinear autocorrelation traces, the spectral phase of the FEL pulse was retrieved based on the EPRIAC algorithm [35]. In this algorithm, the spectral phase is modelled with a Taylor expansion about the carrier frequency $\omega_0$.

$$\varphi(\omega) = \phi_0 + \frac{د\varphi(\omega_0)}{د\omega}(\omega - \omega_0) + \frac{1}{2!}\frac{د^2\varphi(\omega_0)}{د\omega^2}(\omega - \omega_0)^2 + \frac{1}{3!}\frac{د^3\varphi(\omega_0)}{د\omega^3}(\omega - \omega_0)^3 + \cdots.$$

In this study, up to fifth-order terms were taken into account for modelling the FEL pulse. The first two terms represent the absolute phase and the group delay. Since they do not influence the pulse shape or its measurement, these two terms were neglected in the analysis. The second-order, third-order, fourth-order, and fifth-order coefficients were determined by an evolutionary algorithm. First, 200 genes with 4 elements corresponding to each coefficient were randomly generated. Then, the temporal profile of the complex electric field was calculated by Fourier transformation of the spectral amplitude calculated from the LA trace and spectral phase calculated from the coefficients. Then, an FRAC trace could be calculated from the temporal profile of the complex electric field. Here, 200 FRAC traces could be calculated from the 200 genes. The survival parameter of the $i$-th gene was calculated as

$$S_i = \left[\sum_j^N \{FRAC_i(j) - FRAC_{meas}(j)\}^2\right]^{-1},$$

where $FRAC_i(j)$ and $FRAC_{meas}(j)$ are the FRAC traces of the $i$-th gene and the measured result, respectively. A larger value of $S_i$ implies a better fit between two FRAC traces. Using this survival parameter, all genes were ranked, and genes for the next generation were generated based on the evolutionary algorithm. In our analysis, 30 generations were examined for one set of initial conditions, and independent runs were repeated 200 times to avoid being trapped in a local maximum. Thus, 1.2 million genes were examined for one dataset. Finally, the genes that gave the highest survival parameter were selected from all runs and all generations. Then, the pulse structure and phase were calculated from the temporal profile of the complex electric field. Here, we should note that the frequency spectra calculated from the LA traces were filtered by ideal (rectangular) bandpass filters before the EPRIAC analysis and pulse structure retrieval since the LA traces contained widely distributed white noise over the entire frequency range. In the case of analysis of the thermionic operation results, the lower cutoff frequency and the higher cutoff frequency were selected as 23.7 THz and 30.6 THz, respectively, to reduce the influence of white noise. In the analysis of the photocathode operation results, the lower cutoff frequency and the higher cutoff frequency were selected as 20.4 THz and 30.6 THz, respectively.

The software packages for retrieving the spectral phase and calculating the pulse structure and phase were developed in the LabVIEW® environment. To test the software packages, a temporal profile of a complex electric field given by a 1-D numerical simulation of an FEL oscillator with 5% energy extraction efficiency was used. We confirmed that the developed software can retrieve the complex pulse shape and phase expected for a high-efficiency FEL oscillator (see Supplementary Information S1).

*FEL simulation*

Simulations of the lasing behaviour of FEL oscillators are, in general, conducted by one- or three-dimensional codes based on macroparticle tracking [22, 33], in which the electron bunch is divided into many slices along the longitudinal direction and macroparticles are placed in each slice. The motion of the macroparticles is tracked based on the interaction with the optical and undulator fields, and the growth of the radiation field is obtained by averaging the phase-space parameters of the macroparticles over at least one radiation wavelength [49]. This type of simulation is called an averaged code. In the high-efficiency FEL oscillators studied in the present paper, the assumption that the macroparticles are fixed to a slice is not valid because the electron energy significantly changes through the undulator. Therefore, we adopted a one-dimensional unaveraged code, in which macroparticles are not confined to specific slices and may redistribute during propagation. The growth of the radiation field is obtained by superposing the electromagnetic fields created by individual macroparticles [50].

In the simulations shown in Figures 3 and 4, the slippage length, $L_s$, was divided into 624 grid points. We assumed a parabolic electron bunch with a full width of $0.45L_s$ and an FEL gain parameter ρ=0.0039 at the bunch centre for the thermionic cathode mode. For the photocathode mode, we assumed a Gaussian bunch with σ =$0.1L_s$ and ρ=0.0061 at the bunch centre. The round-trip cavity loss was 3%, and the cavity length detuning was altered from $δL$=−$0.0066L_s$ to $δL$ =0 at the 100-th round trip for the thermionic cathode mode and from $δL$=−$0.015L_s$ to $δL$ =0 at the 30-th round trip for the photocathode mode. The number of macroparticles was 1440 for each bunch slice equal to the radiation wavelength.

### Data availability
All the relevant data that support the findings of this study are available from the corresponding authors upon reasonable request.

Supplementary Information for

# Full characterization of few-cycle superradiant pulses generated from a free-electron laser oscillator

Heishun Zen[1*], Ryoichi Hajima[2], Hideaki Ohgaki[1]

[1]Institute of Advanced Energy, Kyoto University, Gokasho Uji, 611-0011 Kyoto, Japan
[2]National Institutes for Quantum Science and Technology, Kizugawa, Kyoto 619-0215, Japan
*e-mail: zen@iae.kyoto-u.ac.jp

## S1. Examination of the developed software with a complex electric field obtained by a one-dimensional numerical simulation of an FEL oscillator with 5% energy extraction efficiency

In this section, we describe the validation process for the software used for the free-electron laser (FEL) pulse measurements. The validation was completed by recovering a known pulse shape generated by a numerical simulation.

Figure S1(a) shows the complex electric field of an FEL pulse obtained by a one-dimensional numerical simulation of the FEL oscillator with 5% energy extraction efficiency and a carrier wavelength of 11.8 μm. The intensity and temporal phase of the FEL pulse are shown in Figure S1(b). This pulse contains multiple pulses and π-phase jumps, similar to the experimental results reported in the main text. From the complex electric field, the linear autocorrelation (LA) trace and the fringe-resolved autocorrelation (FRAC) trace of the FEL pulse were directly calculated and are shown in Figure S2. The obtained spectral distribution without phase information and the FRAC trace were used in the developed software to examine whether the software can accurately retrieve the pulse shape and phase. The original intensity and phase are compared with the retrieved intensity and phase in Figure S3. As can be seen in the figure, the complicated intensity and phase can be accurately retrieved by the developed software. In this examination, the pulse direction was used as the preknown parameter since the FRAC trace does not include information on the pulse front or tail. For actual measurement, a 10-mm-thick ZnSe plate was inserted into the beam path, and information on the pulse shape variation was used to determine the pulse direction, as described in the main text and Supplementary Information S2.

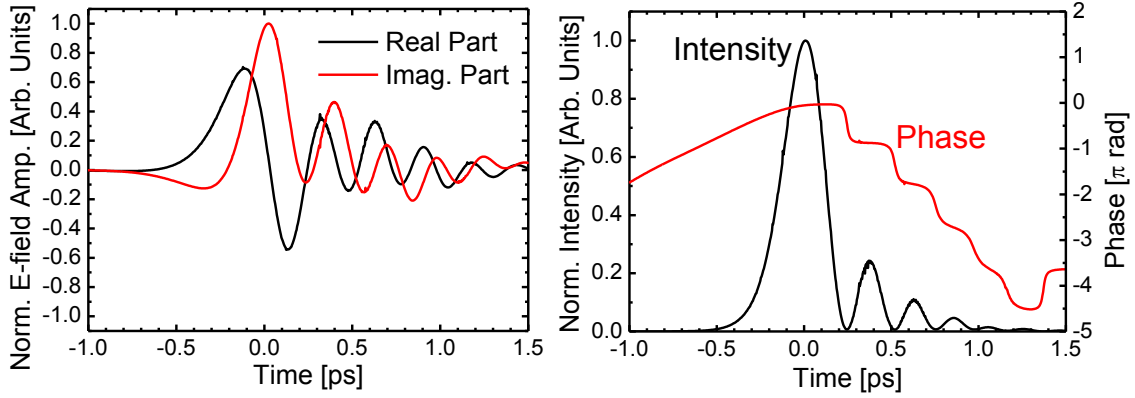

(a)                                    (b)

Figure S1: (a) Complex electric field distribution obtained by a 1D numerical simulation. (b) Intensity and phase distribution calculated from the complex electric field distribution. The phase offset was adjusted to have the maximum phase be zero, and the large linear component of the phase evolution corresponding to the shift of the wavelength from the resonance wavelength used in the numerical simulation and calculation of the complex electric field distribution was also subtracted to display the π-phase jumps.

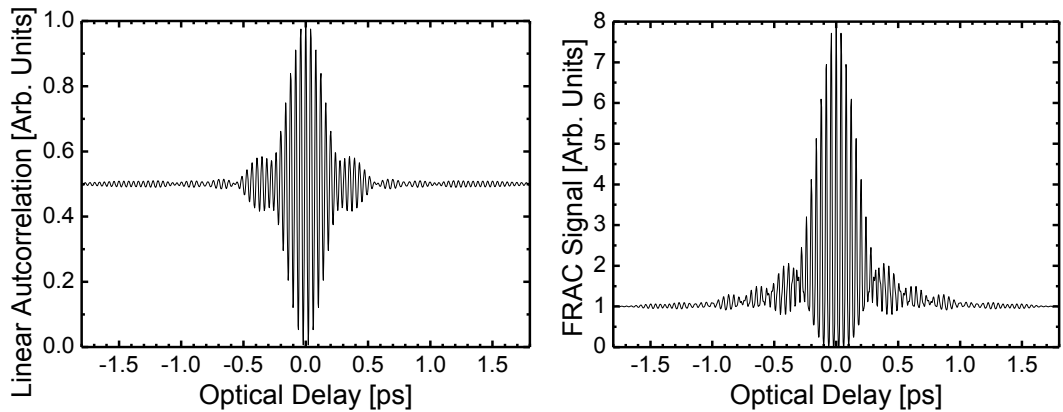

(a)                                    (b)

Figure S2: (a) LA trace and (b) FRAC trace directly calculated from the complex electric field distribution obtained by a 1D FEL simulation.

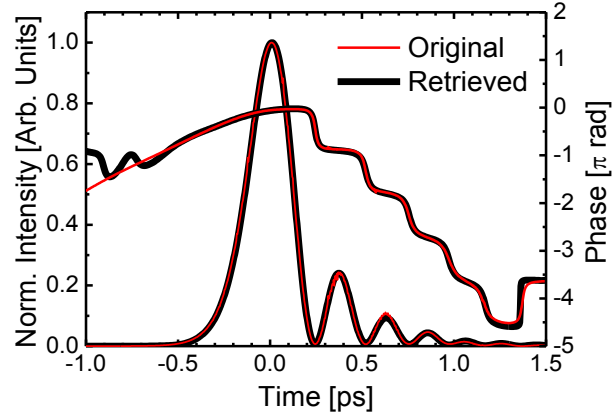

Figure S3: Intensity and phase evolution of the original and retrieved pulses. The time offset and phase offset of the retrieved pulse were adjusted to obtain good matching with the original pulse.

**S2. Determination of the pulse direction by pulse shape measurement with insertion of a dispersive material**

Since the FRAC measurement result has a symmetric distribution over time, the pulse direction cannot be determined only from the FRAC measurement. In this study, we inserted a dispersive material (ZnSe, 10 mm thick, with an antireflection coating on both sides) into the FEL beam path, and the FEL pulse structures with and without the dispersive material were retrieved with the same method as the pulse shape measurement described in the main text. The measured FEL pulses with and without the ZnSe plate are shown in Figure S4. As can be seen in the figure, insertion of the ZnSe plate makes the FEL pulse duration longer. Since the group velocity dispersion of ZnSe at the FEL wavelength (10.3 μm in this experiment) is -1365.9 fs$^2$/mm, the FEL pulse should have a frequency downchirp. The intensity and phase of the FEL pulse without the ZnSe plate are shown in Figure S5. In this case, the temporal phase of the main pulse has a negative second derivative coefficient. This implies that the main pulse has a downchirp. When the pulse direction is flipped, the sign of the temporal phase also flips. In the case of the flipped pulse direction, the measured result has a positive second derivative coefficient, i.e., a frequency upchirp. From the results measured with and without the ZnSe plate, the FEL pulse direction can be determined as the direction shown in Figure S5, i.e., smooth rising with a sech$^2$-like temporal evolution and a successive ringing structure. The pulse direction and chirp direction are consistent with the 1D numerical simulation result (Figure S1(b)).

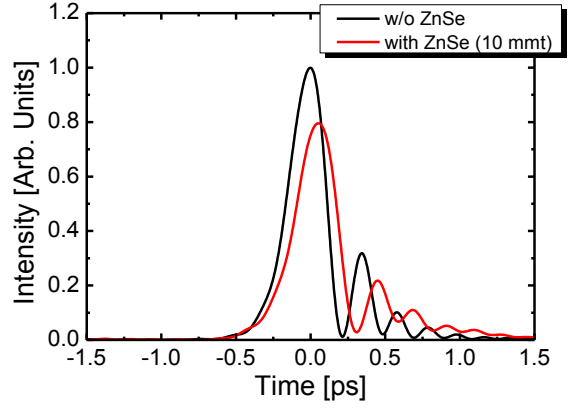

Figure S4: Measured FEL pulses without (black) and with (red) the ZnSe plate in the FEL beam path.

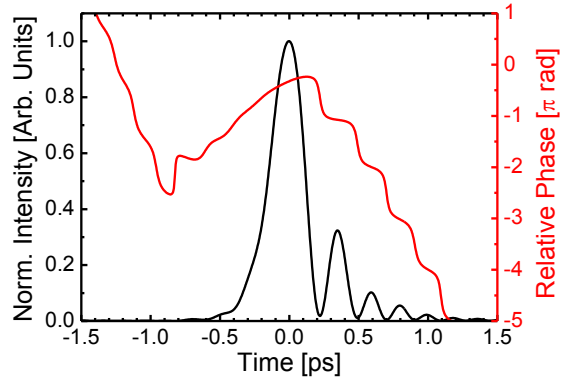

Figure S5: Intensity and phase of the FEL pulse without the ZnSe plate (same as Figure 1 in the main text).

**S3. Retrieval of the FEL pulse structure immediately after the coupling hole**

Since an ultrashort FEL pulse has a quite wide frequency spectrum, as shown in Figure 1(a) and 2(a), the pulse structure strongly depends on the dispersive materials in the FEL beam path. To obtain the original FEL pulse structure immediately after the out-coupling hole of the FEL cavity, the influences of the dispersive materials must be subtracted from the measured pulse structure. Since we can obtain the spectral phase distribution of the FEL pulse by Evolutionary Phase Retrieval from Interferometric AutoCorrelation (EPRIAC) [1] analysis, the FEL pulse structure immediately after the coupling hole can be retrieved by subtracting the phase variation due to the diffractive materials.

The refractive indices of the dispersive components in the beam path were calculated using the Sellmeier equations of the specific materials. In the experiment, there were KRS-5, ZnSe, and KBr windows in the beam path. The Sellmeier equations of these materials are given as [2]

(KRS-5) $n_{\text{KRS}-5}(\lambda)^2 - 1 = \frac{1.8293958\lambda^2}{\lambda^2-0.0225} + \frac{1.6675593\lambda^2}{\lambda^2-0.0625} + \frac{1.1210424\lambda^2}{\lambda^2-0.1225} + \frac{0.04513366\lambda^2}{\lambda^2-0.2025} + \frac{12.380234\lambda^2}{\lambda^2-27089.737^2}$,

(ZnSe) $n_{\text{ZnSe}}(\lambda)^2 - 1 = \frac{4.45813734\lambda^2}{\lambda^2-0.200859853^2} + \frac{0.467216334\lambda^2}{\lambda^2-0.391371166^2} + \frac{2.89566290\lambda^2}{\lambda^2-47.1362108^2}$,

(KBr) $n_{\text{KBr}}(\lambda)^2 - 1 = 0.39408 + \frac{0.79221\lambda^2}{\lambda^2-0.146^2} + \frac{0.01981\lambda^2}{\lambda^2-0.173^2} + \frac{0.15587\lambda^2}{\lambda^2-0.187^2} + \frac{0.17673\lambda^2}{\lambda^2-60.61^2} + \frac{2.06217\lambda^2}{\lambda^2-87.72^2}$,

where $\lambda$ is the wavelength in the unit of micrometres.

The relative difference in the optical phase shift $\Delta\phi_A$ induced by dispersive material A with thickness $L$ at the frequency of interest $f$ from the reference frequency $f_0$ can be calculated as

$$\Delta\phi_A(f - f_0, L) = -2\pi \frac{f n_A\left(\frac{c_0}{f} \times 10^6\right) - f_0 n_A\left(\frac{c_0}{f_0} \times 10^6\right)}{c_0} L$$

because of the theoretical expression of a plane wave in a dispersive medium

$$E(f, z, n) = C\cos\left(2\pi f \left(t - \frac{n}{c_0} z + \phi_0\right)\right),$$

where $z$ and $\phi_0$ are the propagation distance in the dispersive medium and the initial phase before entering the dispersive medium, respectively. This phase shift was calculated for all frequencies and subtracted from the spectral phase distribution obtained by EPRIAC. Then, the pulse shape was obtained by performing complex inverse Fourier transformation using the obtained spectral phase distribution and original intensity distribution. Since the linear component of the spectral phase distribution in the frequency domain, i.e., the group delay, does not change the pulse shape, the linear component was omitted in the calculation results for simplicity.

The above treatment was validated by the experimental results obtained for pulse direction determination shown in Supplementary Information S2. The influence of the 10-mm-thick ZnSe plate was numerically subtracted from the experimental result obtained with the ZnSe plate. The result is shown in Figure S6. As shown in the figure, by subtracting the phase difference distribution due to the ZnSe plate, we can obtain the same pulse shape as that measured without the ZnSe plate.

To retrieve the FEL pulse structure at the out-coupling hole, the phase difference distributions for two KRS-5 plates with a total effective thickness of 7 mm, a ZnSe beam splitter with a total effective thickness of 3.1 mm, and a KBr plate with a thickness of 3 mm were subtracted from the experimental results.

Retrieval of the original pulse shape is important for discussing the detailed characteristics of the FEL pulse and for comparison with the numerical simulation results.

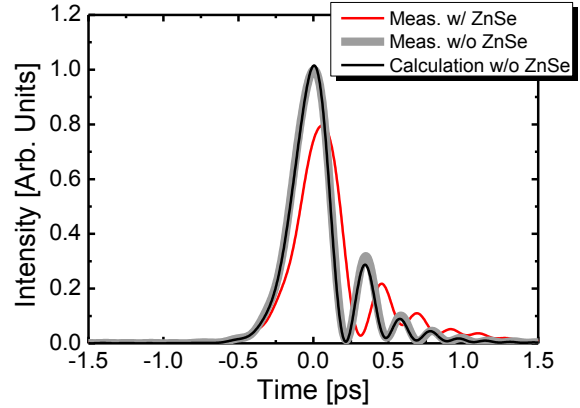

Figure S6: Measured pulse shape with (red) and without (grey) a 10-mm-thick ZnSe plate. The black line represents the pulse shape numerically retrieved by subtracting the influence of the 10-mm-thick ZnSe plate.

## S4. Expected FEL pulse duration with linear chirp compensation and perfect chirp compensation under photocathode operation

As shown in Figure 3 in the main text, the superradiance FEL pulses inherently have a frequency downchirp. The FEL pulse duration can be further shortened by chirp compensation. Linear and perfect chirp compensation can be numerically performed since the spectral intensity and phase were obtained by the measurement. The linear and perfect chirp compensation results are shown in Figure S7. In the case of linear chirp compensation, the pulse duration of the main pulse can be shortened down to 121 fs in full width at half maximum (FWHM) (3.4 optical cycles at 28 THz). Even with linear chirp compensation, ringing remains after the main pulse. When the chirp is perfectly compensated, the ringing disappears, and the pulse duration is shortened down to 105 fs in FWHM (2.9 optical cycles at 28 THz).

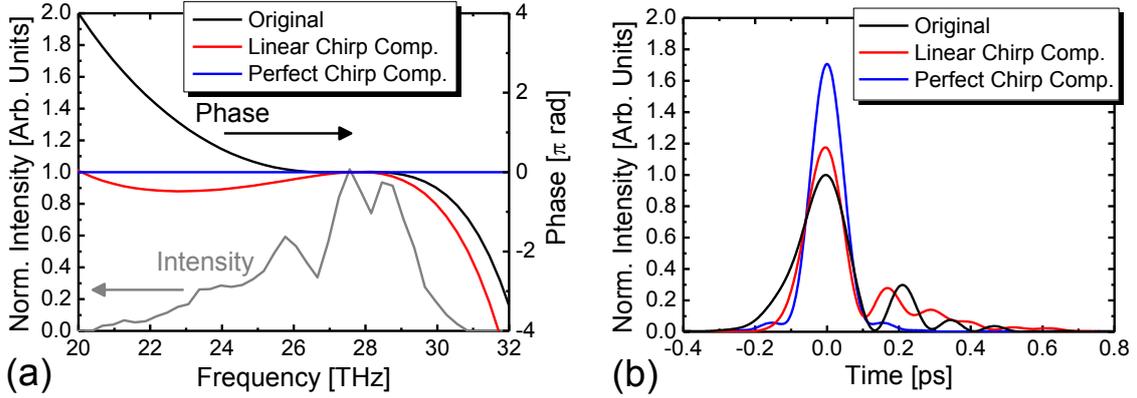

Figure S7: Result of linear and perfect chirp compensation together with the original distribution. (a) Spectral intensity and phase distribution. (b) Pulse structure.

**S5. One-dimensional FEL simulation code**

In this section, we describe details of the simulation code used in the present study and calculation results that could not be included in the main text.

The generation of coherent radiation in an FEL can be calculated by numerically solving the Maxwell equation coupled with equations to derive the motion of electrons in the radiation and undulator magnetic fields. Since the number of electrons in a bunch, ~$10^9$ for a 200 pC bunch, is too large, so-called macroparticles are introduced to conduct simulations with reasonable computer resources. In general FEL simulation codes, the electron bunch is divided into many slices along the longitudinal axis, and macroparticles are prepared in each slice. The motion of the macroparticles within a bunch slice is then tracked by considering the energy exchange between the electron and the radiation field. The evolution of the radiation electric field is calculated from the local driving source, i.e., the Fourier component of the beam current corresponding to the radiation wavelength. The Fourier component is obtained by averaging the contribution of macroparticles over at least one radiation wavelength. The simulation is, therefore, called the averaged code.

Averaged simulations, both one-dimensional and three-dimensional, have been widely used in the analysis of FELs. The averaged code is, however, not appropriate for the analysis of infrared FEL oscillators with large extraction efficiency because the assumption that the macroparticles are fixed to a slice is not valid for a high-efficiency FEL oscillator, in which some of the electrons significantly change energy and move across bunch slices through the undulator. Therefore, we adopted a one-dimensional unaveraged code, in which macroparticles are not confined to specific slices and may redistribute during propagation.

Following the paper that proposed the unaveraged FEL simulation [3], we modified our one-dimensional simulation code previously developed for analysis of FEL oscillators [4]. In the simulation code, the longitudinal variable is normalized as $\zeta=(z-ct)/L_s$, the time for traversing the undulator is normalized as $\tau= ct/L_u$, and the electron energy is normalized as $\mu=(\Delta E-E_0)/4\pi N_u$, where $N_u$ is the number of undulator periods, $L_s=N_u\lambda$ is the slippage distance, and $\Delta E$ is the electron energy deviation from the resonance energy $E_0$. Macroparticles are prepared along the longitudinal coordinate $\zeta$ by reflecting the electron bunch profile, and the particles can move freely along $\zeta$ while passing through an undulator. The growth of the radiation field is updated every time step by superposing the electromagnetic fields created by individual macroparticles.

As discussed in paper [3], the effects of coherent spontaneous emission (CSE) appear explicitly in the unaveraged simulations. The CSE has another impact on the simulations of the high-efficiency FEL oscillator because the CSE increases the amplitude of the effective shot noise, which plays an essential role in both initiating the lasing and maintaining the lasing after saturation at the perfectly synchronized cavity length [5].

Figures S8 and S9 show the results of the unaveraged simulation with the same parameters as those used for the simulations of the photocathode mode operation presented in the main text. In these figures, the evolution of the FEL macropulse and pulse shapes for the 10[th], 50[th], 100[th] and 250[th] round trips are plotted. The normalized radiation intensity, $|A|^2$, is defined such that $\rho|A|^2$ gives the ratio between the energy density of the radiation and the resonant electron beam [6]. In Figure S9, continuous narrowing of the main pulse and the superluminal property can be identified. The distribution of the macroparticles in the longitudinal phase space at the end of the 210[th] round trip is plotted in Figure S10, which depicts deformation of the bunch envelope due to the large drift of macroparticles in addition to the formation of microbunches whose interval is equal to the radiation wavelength.

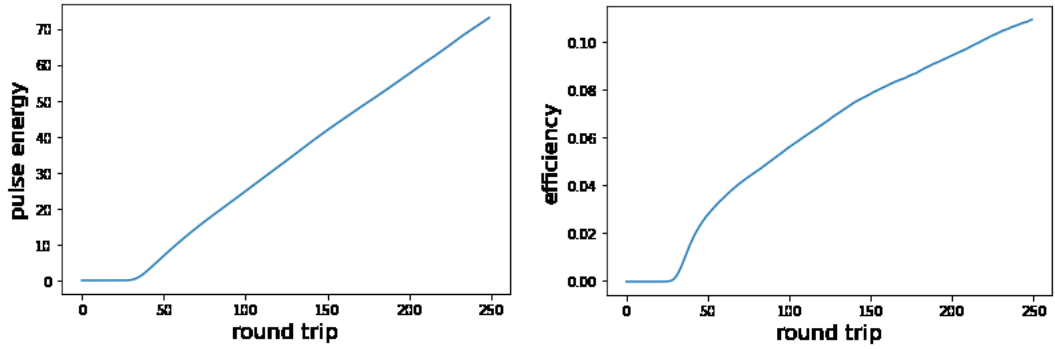

(A)  (B)

Figure S8: Evolution of a macropulse calculated for photocathode mode operation. (A) FEL pulse energy and (B) extraction efficiency as a function of the round trip number.

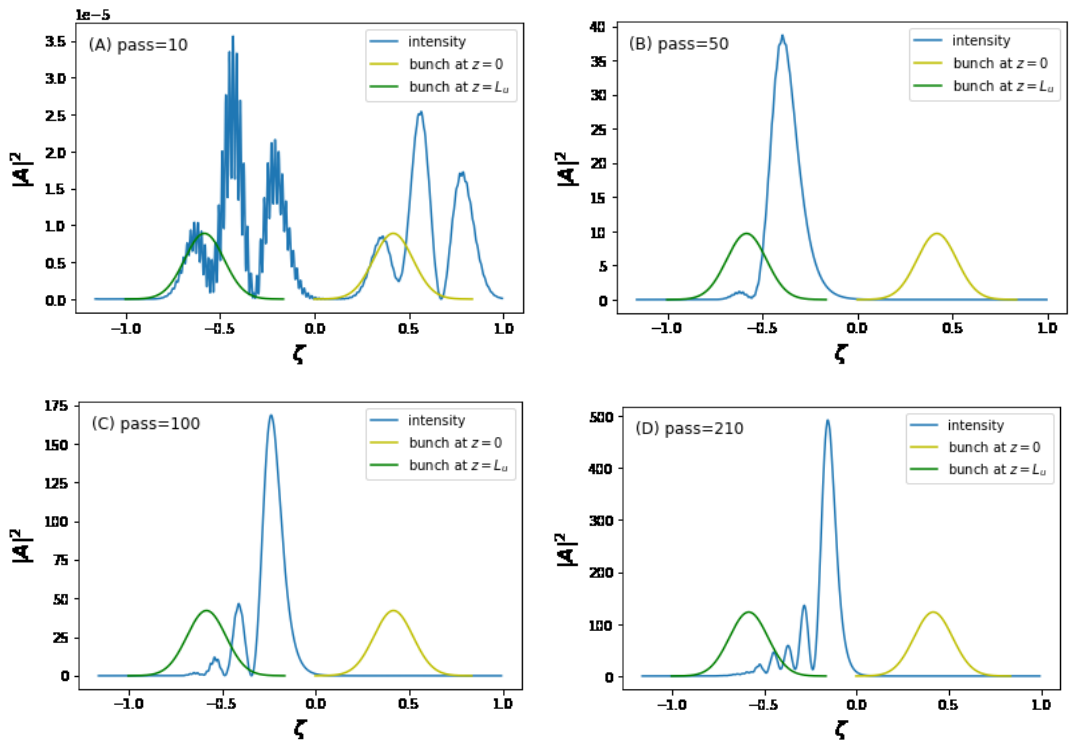

Figure S9: FEL pulse shapes for the (A) 10th, (B) 50th, (C) 100th and (D) 250th round trips calculated by the unaveraged code.

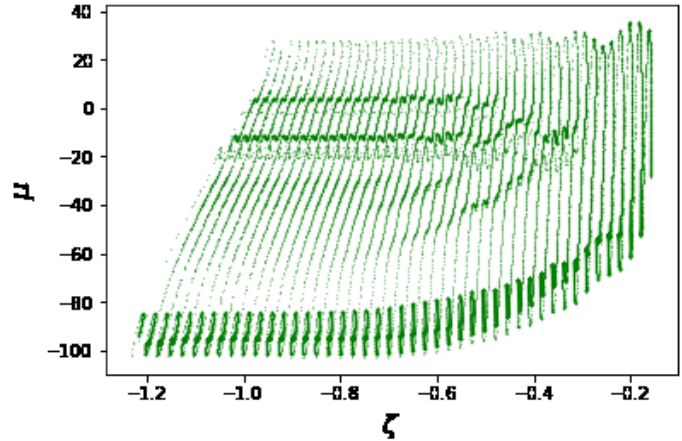

Figure S10: Distribution of the macroparticles at the end of the 210th round trip.

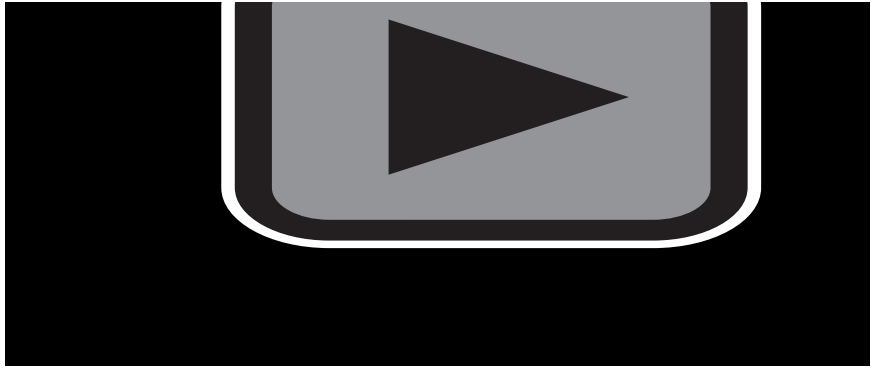

Supplementary Video 1: A movie to represent the phase space motion of macroparticles in a simulation for the KU-FEL experiment with the photocathode mode.
(Left panel): The dots are the macroparticles and the red line is the radiation field, which is defined so that the positive is the acceleration field. Local changes in the macroparticle energy are plotted as the blue bars with units at the right axes.
(Right panel): The location of the macroparticles with respect to the FEL pulse is plotted by the red line.